\documentclass[12pt,preprint]{aastex}

\newcommand{\HI}{H\kern0.1em{\sc i}}
\newcommand{\unm}{1}
\newcommand{\kipac}{2}
\newcommand{\haystack}{3}
\newcommand{\grb}{GRB\,030329}
\newcommand{\simlt}{\mathrel{\hbox{\rlap{\hbox{\lower4pt\hbox{$\sim$}}}\hbox{$<$}}}}
\newcommand{\simgt}{\mathrel{\hbox{\rlap{\hbox{\lower4pt\hbox{$\sim$}}}\hbox{$>$}}}}

\shorttitle{VLBI observations of GRB\;030329}
\shortauthors{Pihlstr\"om et al.}

\begin{document}

\title{Stirring the Embers: High Sensitivity VLBI Observations of GRB~030329}

\author{Y.M. Pihlstr\"om\altaffilmark{\unm}, G.B. Taylor\altaffilmark{\unm}, 
J. Granot\altaffilmark{\kipac}, \& S. Doeleman\altaffilmark{\haystack} }

\altaffiltext{\unm}{University of New Mexico, Dept. of Physics and
Astronomy, Albuquerque, NM 87131, USA; ylva,gbtaylor@unm.edu}

\altaffiltext{\kipac}{KIPAC, Stanford University, P.O. Box 20450, MS
29, Stanford, CA 94309; granot@slac.stanford.edu}

\altaffiltext{\haystack}{MIT Haystack Observatory, Off Route 40,
  Westford, MA 01886; dole@haystack.mit.edu}

\begin{abstract}
We present high sensitivity Very Long Baseline Interferometry (VLBI)
observations 806 days after the $\gamma$-ray burst of 2003 March 29
(GRB~030329).  The angular diameter of the radio afterglow is measured
to be $0.347 \pm 0.09\;$mas, corresponding to $0.99 \pm 0.26\;$pc at
the redshift of GRB~030329 ($z = 0.1685$). The evolution of the image
size favors a uniform external density over an $R^{-2}$ wind-like
density profile (at distances of $R \gtrsim 10^{18}\;$cm from the
source), although the latter cannot be ruled out yet. The current
apparent expansion velocity of the image size is only mildly
relativistic, suggesting a non-relativistic transition time of $t_{\rm
NR} \sim 1\;$yr. A rebrightening, or at least a significant flattening
in the flux decay, is expected within the next several years as the
counter-jet becomes visible (this has not yet been observed). An upper
limit of $<1.9\;c$ is set on the proper motion of the flux centroid.
\end{abstract}


\section{Introduction}

Our understanding of the origin of gamma-ray bursts (GRBs) has
advanced rapidly in recent years since the first X-ray \citep{cos97},
optical \citep{van97} and radio \citep{fra97} afterglows were
discovered. We now know that a GRB represents the quick ($\lesssim
10^2\;$s) release of a large amount of energy -- about $10^{51}\;$erg,
when corrected for jet collimation \citep{Frail01,BFK03}. The prompt
$\gamma$-ray emission is usually attributed to internal shocks due to
variability within the ultra-relativistic outflow (with a Lorentz
factor $\gtrsim 10^2$), while the afterglow is from a strong
relativistic shock that is driven into the external medium as it
decelerates the ejecta \citep[for a review see][]{Piran05}. The
afterglow emission is predominantly synchrotron radiation, and its
spectral and temporal evolution \citep{SPN98,GS02} are governed by
such factors as the structure and dynamical evolution of the
relativistic jet \citep[for a review see][]{Granot06}, and the
environment \citep[e.g., constant density versus a wind-like density
profile;][]{cl00} which are not well constrained.

GRB~030329, discovered by HETE-2 \citep{van03}, and localized rapidly
in the optical bands by Peterson \& Price (2003) \nocite{pet03}
represents a unique opportunity for VLBI observations. At a redshift
of $z=0.1685$ \citep{gre03} this is the second nearest cosmological
burst in the northern sky detected to date \citep[only GRB~060218 is
nearer at a redshift $z=0.0331$,][]{mirabal06}. Observation with the
VLA show the \grb\ radio afterglow to be the brightest detected to
date, with a maximum flux density of 55 mJy at 43 GHz one week after
the burst \citep{bkp+03}.

Most VLBA studies so far have been restricted to rather distant GRBs
($z\sim 1$).  While these observations have proved to be invaluable in
studying the flux evolution of the afterglow and providing accurate
positions \citep[in one case localizing the GRB within 0.3 kpc of the
center of its radio host galaxy;][]{BKF01}, in all cases the afterglow
faded before it could be resolved by the VLBA.  In the case of
GRB~030329 we have an exceptional instance of a radio afterglow near
and bright enough that it can be resolved with VLBI. Its exceptional
brightness enables the radio afterglow to be studied for much longer
times than has been typical.

The VLBA campaign on GRB 030329 \citep{tay04,tay05} resolved the
afterglow image and showed an increase in its diameter from $65 \pm
22\;\mu$as at $t = 25\;$days after the burst to $172 \pm 43\;\mu$as at
$t = 83\;$days.  Only slow growth was seen out to $t = 217\;$days when
the diameter was measured to be $176 \pm 80\;\mu$as, possibly
indicating a transition to non-relativistic expansion.  Here we
present very late time observations of GRB 030329 that were performed
at $t = 806\;$days after the burst with the goal of differentiating
between afterglow models, and constraining the jet structure and
dynamics as well as the environment.  Such observations can also test
the hypothesis of a double-sided jet, which would imply a significant
contribution from the counter-jet as the flow becomes sub-relativistic
at late times.

Assuming a $\Lambda$ cosmology with $H_0 = 71\;{\rm
  km\;s^{-1}\;Mpc^{-1}}$, $\Omega_M = 0.27$ and $\Omega_\Lambda=0.73$,
the angular-diameter distance of \grb\ at $z=0.1685$ is
$d_A=587\;$Mpc, and 1 milliarcsec corresponds to 2.85 pc.

\section{Observations}\label{sec:obs}

\subsection{Late Time VLBI Observations}
Late time observations require high sensitivity to detect the fading
afterglow. To achieve this high sensitivity we employed the largest
aperture radio telescopes available, and used a bandwidth of $64\;$MHz
in two circular polarizations. The VLBI observations were taken at
$5\;$GHz on 2005 June 12, $806\;$days after the burst, with a global
array including the $100\;$m Green Bank Telescope (GBT) of the
NRAO\footnote{The National Radio Astronomy Observatory is operated by
  Associated Universities, Inc., under cooperative agreement with the
  National Science Foundation.}, the Effelsberg $100\;$m
telescope\footnote{The $100\;$m telescope at Effelsberg is operated by
  the Max-Planck-Institut f{\"u}r Radioastronomie in Bonn.}, the
$305\;$m Arecibo telescope\footnote{The Arecibo Observatory is part of
  the National Astronomy and Ionosphere Center, which is operated by
  Cornell University under a cooperative agreement with the National
  Science Foundation.}, the Westerbork (WSRT) tied array\footnote{The
  Westerbork Synthesis Radio Telescope is operated by the ASTRON
  (Netherlands Foundation for Research in Astronomy) with support from
  the Netherlands Foundation for Scientific Res earch (NWO)}, and the
$25\;$m MarkII telescope of Jodrell Bank\footnote{Jodrell Banks
  Observatory's VLBI participation is hosted by the national Facility
  of the Particle Physics and Astronomy Research Council (PPPARC)}
(replaced from the initially scheduled Lovell telescope due to the
fast source-switching phase-reference schedule). In all, the combined
effective collecting area is $0.097\;{\rm km^2}$. The on-source time
on GRB~030329 was $1.6\;$hr for all telescopes except for Arecibo
($0.9\;$hr). Data were recorded at $1\;{\rm Gb\;s^{-1}}$ and 2-bit
sampling at all stations except for Green Bank, where data were
recorded at $512\;{\rm Mb\;s^{-1}}$ and 1-bit sampling. The correlator
saw the Green Bank data as 2-bit data, with the magnitude bit always
set to high. The post-correlation 2-bit van Vleck correction
thereafter processed the data as usual. The observations were
correlated at the Joint Institute for VLBI in Europe (JIVE).

The nearby (0.8$^\circ$) source J1048+2115 was used as the
phase-reference calibrator, and was observed regularly in a cycle of 2
minutes on the target and 1 minute on the calibrator. Since J1048+2115
is not tied to the International Celestial Reference Frame (ICRF), the
ICRF source J1051+2119 was observed every 30 minutes. This allowed an
accurate determination of the position of the phase-reference source
and subsequently the target. Self-calibration solutions for the
amplitude were transferred from the stronger J1051+2119 calibrator,
which corrected for a large amplitude offset affecting all the left
hand circular polarization IFs of the MarkII telescope. In proposing
these observations, we aimed for a theoretical noise of around 6
$\mu$Jy/beam. However, a number of factors resulted in a decreased
sensitivity in our final map. These factors includes the replacement
of Lovell with the smaller MarkII telescope, and missing or bad
data. As a result, a noise of 15 $\mu$Jy/beam is reached in the final
image.

\subsection{VLA Observations}

Multi-frequency VLA observations of \grb\ have been presented
\citep{bkp+03, fra05} between $1\;$day and $1\;$yr after the burst.
Here we present VLA observations at 5\,GHz taken from the NRAO archive
between 380 and 740 days after the burst (see Table 1 for more
details). The VLA data were reduced in AIPS following standard
procedures \citep{tay99}. The nearby (1.6$^\circ$ separation)
calibrator J1051+213 was used to determine the complex antenna gains.
Absolute flux density calibration was tied to short observations of
3C286.

Using this VLA data in combination with published WSRT data
\citep{vanderhorst05}, we find that the fall in flux densities from
day 59 to day 806 (Fig.~1) are well described by a power law, $F_\nu
\propto t^{-\alpha}$, with a slope $\alpha = 1.23\pm 0.03$. In the end
of 2009, the data recording rate of the VLBA is planned to increase to
$4\;{\rm Gb\;s^{-1}}$ compared to its current limit of $0.5\;{\rm
  Gb\;s^{-1}}$. The inclusion of the VLBA at this recording rate, as
well as the large antennas used for this experiment, should provide an
increase in sensitivity by a factor $\sim 3$, or a theoretical rms
noise of $\sim 2\;\mu$Jy. Assuming that \grb\ continues to fade in the
same way, then we expect a flux density in the end of 2009 (6.6 years
after the burst) of $\sim 70\;\mu$Jy. Thus it should be quite
practical to follow the growth of \grb\ for at least another 3-5
years.

\section{Results}

We fit a symmetric, two-dimensional circular Gaussian to the measured
visibilities on \grb\ and find a full-width at half maximum (FWHM)
size of $0.347 \pm 0.09\;$mas.  As in \cite{tay04} the error on the
size is estimated from signal-to-noise ratios and from Monte-Carlo
simulations of the data using identical ($u$,$v$) coverage, similar
noise properties, and a Gaussian component of known size added.  The
standard deviation of the recovered sizes, model-fitted in the same
way as we treat the observations, was found to be $0.07\;$mas.  Given
two similar estimates of the error, we adopt the more conservative
limit of 0.09 mas based on signal-to-noise ratios and the resolution
of the observations.

We also obtain a position for \grb\ of R.A.  10$^{\rm h}$44$^{\rm
  m}$49.95957$^{\rm s}$ and Dec. 21$^\circ$31'17.4375'' with an
uncertainty of $0.2\;$mas in each coordinate.

\section{Discussion}

\subsection{Angular Size Measurements}\label{sec:size}

The entire history of expansion for \grb\ is shown in
Fig.~\ref{growth}.  The first measurement at 15 days comes from a
model-dependent estimate of the quenching of the scintillation
\citep{bkp+03}.  The uncertainties on this size estimate are large due
to the dependence of the measurement on estimated properties of the
interstellar medium along this particular line-of-sight. Furthermore,
a limb-brightened image has a smaller effective area (similar to a
thin ring) compared to a uniform image of the same diameter. This
reduces the effective number of regions within the image with
independent brightness variations for a given image diameter. Thus, in
a limb-brightened image (as expected here), the image diameter might
be under-estimated by a factor larger than that for the fit to a
circular Gaussian that is used in the other epochs (see discussion
below Eq.~\ref{beta_app_av}), thus potentially causing a systematic
under-estimate of the source size at $15\;$days, compared to the other
epochs that are shown in Fig.~\ref{growth}.

Our directly measured size of $0.347 \pm 0.09\;$mas at $806\;$days is
a factor of $\sim 2$ larger than the size of $0.176 \pm 0.08\;$mas
measured at $217\;$days \citep{tay05}, which was quite close to the
size of $0.172 \pm 0.043\;$mas measured at $84\;$days
\citep{tay04}. We consider the possibility of no expansion between
days 217 and 806, which at a first look at Fig.~\ref{growth} might be
a plausibility. However, the points at 217 and 806 days are separated
by $2\sigma$, and the edges of the $1\sigma$ regions just touch each
other. At the other extreme, the size could have changed with a factor
4.6. In this case, the slope will be inconsistent with the earlier
point at day 84, unless the source increased its expansion speed or
shrank in size. We find this implausible, and that an expansion equal
to (or slower than) a factor of $\sim 2$ is more likely.

Adopting measured size increase of a factor of $\sim 2$, this implies
a continued growth of the image size, and confirms a gradual decrease
in its apparent expansion speed, $\beta_{\rm app}c$. In Fig.~3 we show
the decline of $\langle\beta_{\rm app}\rangle$ with time, calculated
from the measured sizes using
\begin{equation}\label{beta_app_av}
\langle \beta_{\rm app}\rangle =
\frac{(1+z)R_\perp}{ct} = \frac{\theta_R d_M}{ct}\ ,
\end{equation}
where $\theta_R = R_\perp/d_A = (1+z)R_\perp/d_M$ and $R_\perp$ are
the angular and physical radius of the image, respectively, and $d_M$
is the proper distance to the source. We note that similar to
arguments for no expansion in Fig.~2, it could be argued that the
apparent expansion speed (Fig.~3) at later times is constant. This
would mean that the source size grows linearly with time (i.e.\ a
factor of 3.7 growth between day 217 and 806). As mentioned above, this
type of growth requires a rather strange behavior given the point at
day 84. Thus, a decrease in the apparent expansion speed seems more
reasonable.

The values for the apparent size depend somewhat upon the intrinsic
surface brightness profile, which for circular Gaussian, a uniform
disk, and a ring, vary roughly as 1:1.6:1.1 \citep{tay04}.  In
practice, the image is limb-brightened \citep{gps99a,gl01}, and
perhaps slightly closer to a thin ring than to a uniform disk,
suggesting a correction factor of $\sim 1.4$. This correction factor
might change with time, since the surface brightness profile within
the afterglow image is expected to change across the non-relativistic
transition, and possibly also during the post-jet break relativistic
stage. However, the afterglow image during both of these stages has
not been studied yet in sufficient detail to provide a reliable
description of the exact temporal evolution of this correction
factor. At any rate, this effect is expected to be fairly mild, and
smaller than the factor of $\sim 2$ growth in the image size between
$217\;$days and $806\;$days.

Taking this correction factor into account, we find $\langle
\beta_{\rm app}\rangle \approx 1.2$ after $806\;$days, i.e. that the
current expansion velocity (which is expected to be somewhat smaller
than the average value $\langle \beta_{\rm app}\rangle$) is only
mildly relativistic, while at earlier times it was considerably higher
and more relativistic, as implied by the value of $\langle \beta_{\rm
  app}\rangle \approx 8$ after $25\;$days. The non-relativistic
transition occurs when $\beta_{\rm app} \approx 1$, i.e. when $\langle
\beta_{\rm app}\rangle \sim 2$, the exact value depending on the
detailed expansion history \citep{gra05}. This suggests a
non-relativistic transition time of about $t_{\rm NR} \sim
1\;$yr. Based on a joint temporal and spectral fit to the light curve,
\citet{fra05} find $t_{\rm NR} \sim 50\;$days, and \citet{resmi05} use
simultaneous radio, millimeter and optical observations to derive
model-dependent non-relativistic transition times of $t_{\rm NR} \sim
42$ and $\sim 63\;$days. Moreover, multi-frequency radio data
presented by \cite{vanderhorst05} is fitted with a $t_{\rm NR} \sim
80\;$days. All these transition times are condsiderably shorter than
our estimate. In principle their models could be refined using the
direct size estimates presented here to better constrain the
energetics of the burst and other physical parameters of the models.

\begin{figure}
\plotone{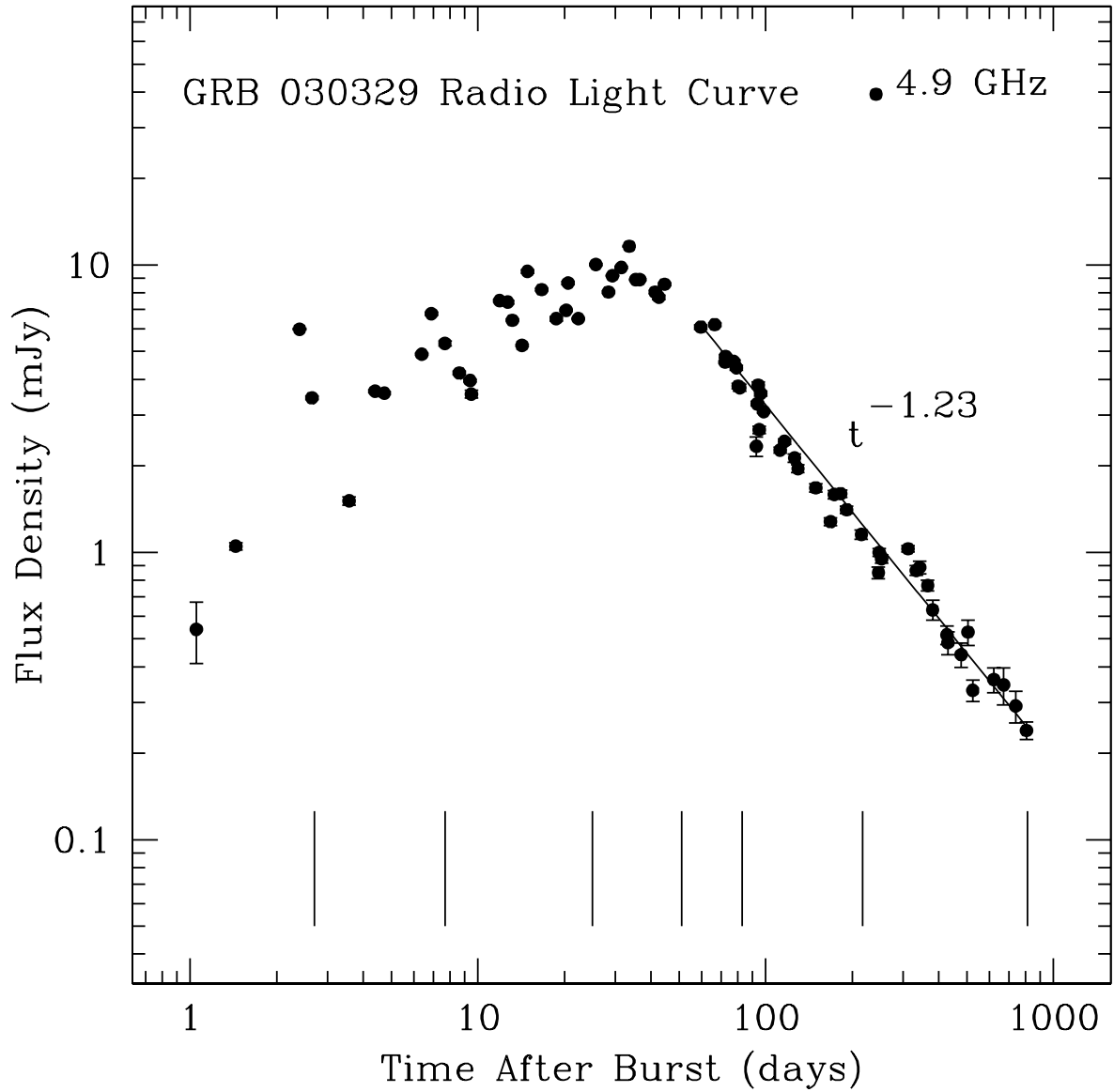}
\caption{The $4.9\;$GHz light curve for GRB~030329 from
publicly available data and these observations.  The VLBI epochs are
indicated by vertical lines between $3\;$days and $806\;$days after the
burst. The data from day 59 to day 806 are well described by
a power law, $F_\nu \propto t^{-1.23\pm 0.03}$ ({\it solid line}).
\label{FIG1}}
\end{figure}

\begin{figure}
\plotone{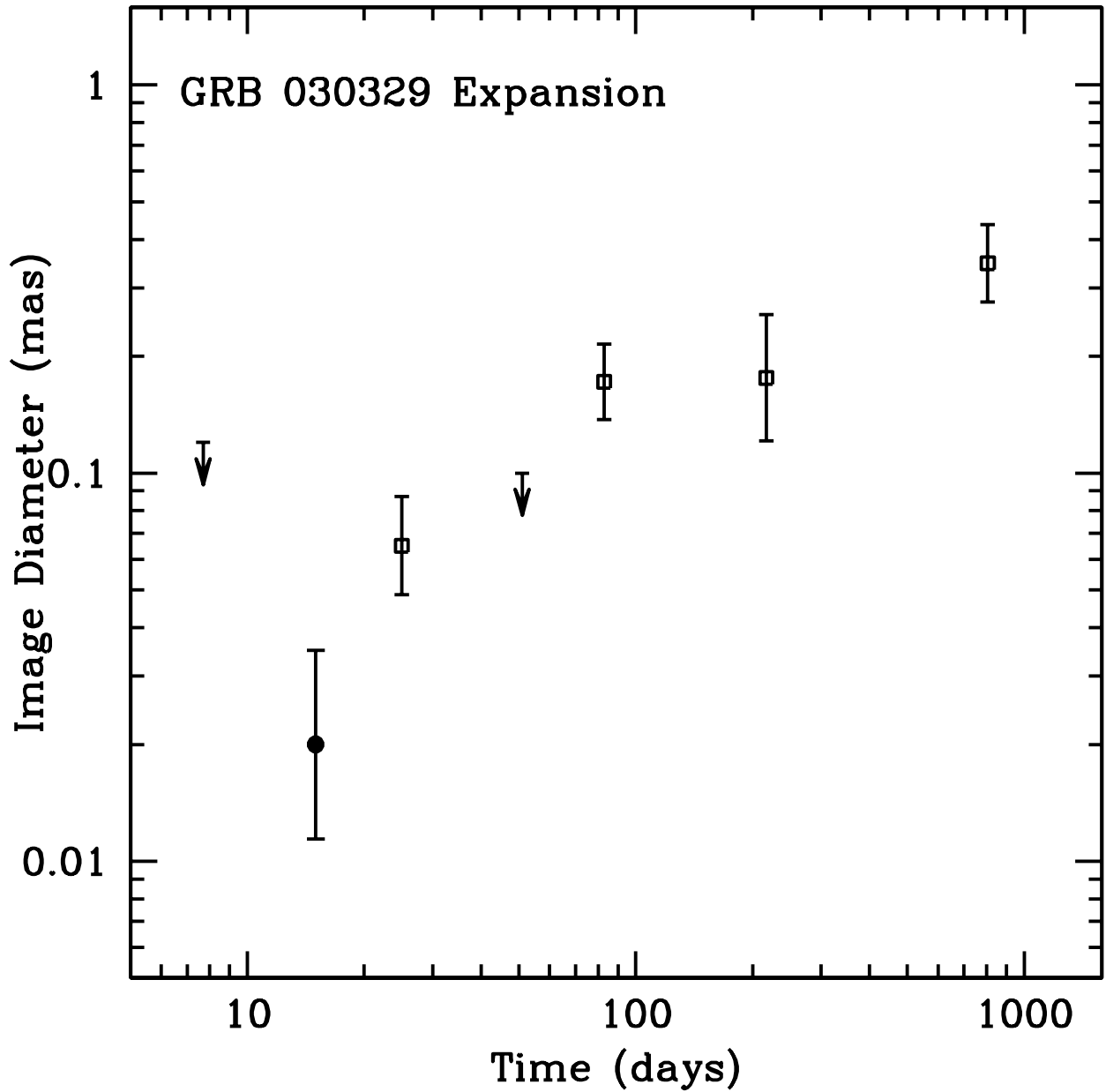}
\caption{ The apparent expansion
of GRB~030329 derived from measurements and limits on the angular size
as a function of time.  The two upper limits at $7.7\;$days and
$51\;$days are from \citet{tay04}, as are the measurements on days 25
and 83 (open squares). The measurement on day 217 is from
\citet{tay05} (open square). The measurement on day 15 (filled
circle) is a model dependent estimate based on the quenching of
scintillation by \citet{bkp+03}. Finally, the measurement on day 806
(open square) comes from this work. The {\it lower panel} shows the
evolution of the average apparent expansion velocity.
\label{growth}}
\end{figure}

Radio re-brightening of \grb\ has been predicted by \cite{gl03} as the
counter jet becomes non-relativistic and its emission is no longer
strongly beamed away from us. \cite{li04} estimated a level of
$0.6\;$mJy at $15\;$GHz, $1.7\;$yr after the burst, due to this
effect.  The actual flux density around that time, however, was $\sim
3.3$ times lower than their prediction ($0.36\;$mJy at $4.86\;$GHz
after $621\;$days, with $F_\nu \propto \nu^{-0.6}$ in the radio,
between the two frequencies, implying $\approx 0.18\;$mJy at
$15\;$GHz). More importantly, we find that the late time $4.9\;$GHz
light curve, from $59\;$days and up to $806\;$days, is consistent with
a single power law decay of $F_\nu \propto t^{-1.23\pm 0.03}$ (see
Fig.~\ref{FIG1}), which implies that up to $\approx 800\;$days after
\grb\ there is still no significant contribution to the observed flux
from the counter jet. Light travel effects might cause the
contribution from the counter jet to become significant at a somewhat
later time, up to a factor of several after $t_{\rm NR}$ when
$\beta_{\rm app} \approx 1$, i.e. up to several years after the burst
\citep[this effect was taken into account by][but they estimated a
somewhat lower value for $t_{\rm NR}$ of $\sim
120\;$days]{li04}. Alternatively, if the counter jet encounters a
somewhat lower external density than the forward jet, it would become
non-relativistic at a later lab frame time, and light travel effects
would cause its contribution to the observed flux density to become
significant at an even later time (as well as being intrinsically less
luminous than the forward jet at the same radius). If such a
re-brightening (or at least a significantly shallower flux decay)
eventually occurs within the next several years, it may enable
measurement of the image size for a longer time.

Fig.~\ref{FIG4} demonstrates how the measured evolution of the image
size for \grb\ compares with the predictions of different theoretical
models. The data constrain the external density profile at radii $R
\gtrsim 10^{18}\;$cm, and favor a uniform density external medium
($\rho_{\rm ext} = Ar^{-k}$ with $k = 0$) over a stellar wind
environment \citep[$k=2$, as expected for a massive star
progenitor;][]{cl00}, although it is hard to rule out the latter due
to the sizeable errors on the source size, and the theoretical
uncertainty in the jet structure and dynamics. The degree of lateral
spreading of the jet is harder to constrain, since its effect on the
evolution of the observed image size around the times when it was
measured is somewhat smaller than that of the external density profile
(see Fig.~\ref{FIG4}).

\subsection{Constraints on Proper motions}

Solving for proper motion using all the VLBI observations to date, we
derive $\mu_{\rm r.a.}=0.12\pm 0.15$ mas yr$^{-1}$ and $\mu_{\rm
dec.}=-0.11\pm 0.08$ mas yr$^{-1}$, or an angular displacement over
$806\;$days of $0.36 \pm 0.38\;$mas (Fig.~5). These observations are
consistent with those reported by \cite{tay04,tay05}, and impose an
even stronger limit on the proper motion. The implied limit on the
proper motion in the plane of the sky is $< 0.38\;$mas (corresponding
to $< 1.08\;$pc) in $806(1+z)^{-1} \approx 690\;$days, or $< 1.87\;c$
($2\;\sigma$). 

Our new upper limits on the proper motion are about a factor of 2
better than previous limits. The proper motion of the flux centroid is
constrained to be less than the current diameter of the image, which
is consistent with the expectations for a narrow double sided
jet,\footnote{The observed image angular diameter around $t_{\rm NR}$
  is expected to be $\gtrsim 2\theta_0R(t_{\rm NR})/d_A =
  2\theta_0\theta_{\rm NR}$, which is a factor of $\gtrsim
  2\theta_0/\theta_{\rm obs} > 2$ larger than the expected maximal
  angular displacement of the flux centroid.}  where the maximal
angular displacement of the flux centroid is $\lesssim \theta_{\rm
  obs}\theta_{\rm NR}$ and is obtained around $t_{\rm NR}$, where
$\theta_{\rm NR} = R(t_{\rm NR})/d_A$ \citep{gl03}, since for \grb\
the viewing angle is within the jet, $\theta_{\rm obs} <
\theta_0$. This consistency is independent of the value of the jet
initial half-opening angle, $\theta_0$, which is somewhat
controversial \citep{gnp03,bkp+03,she03,png04,granot05,gorosabel06}.



\section{Conclusions}

We present measurements of the $5\;$GHz flux density and size of the
radio afterglow from \grb\ over a period of 2.2 years.  These
observations clearly demonstrate that the expansion velocity has
slowed down over time, with a transition to the non-relativistic
regime at $\sim$1 yr. The evolution of the afterglow image size favors
a uniform external density over a wind-like stratified external medium
(at $R \gtrsim 10^{18}\;$cm), although the data still do not enable to
conclusively rule out a wind-like density profile (see
Fig.~\ref{FIG4}). We also present a tight upper limit on the apparent
motion of the flux centroid ($< 1.87\;c$, a factor of 2 better than
previous limits), which are consistent with the predictions of
standard afterglow theory.

Although GRB~030329 has faded considerably, it is likely that VLBI
observations will continue to be useful for determining expansion size
estimates in the future. With VLBI recording rates expected to
increase to $4\;{\rm Gb\;s^{-1}}$ over the next few years, thermal
noise levels of the VLBA at 5\,GHz should decrease to $\sim 8.5\;\mu$Jy,
and arrays of larger antennas, such as the one used in this work, will
reach theoretical thermal noise levels of $\sim 2\;\mu$Jy. This
anticipated increase in VLBI capability should keep pace with the flux
density of GRB~030329 for the next few years. A rebrightening, or at
least a significantly shallower flux decay, is expected within the
next several years, as the counter jet begins to significantly
contribute to the observed flux. Further VLBI observations should
improve constraints on the density profile of the external medium into
which the afterglow shock is expanding, and should eventually reveal
the shape of the afterglow.

\acknowledgments

This research has made use of NASA's Astrophysics Data System. The
research of JG is supported by the US Department of Energy under
contract DEAC03-76SF00515.

\begin{deluxetable}{lrrrrc}
\tablecolumns{7}
\tablewidth{0pt}
\tablecaption{Observational Summary\label{Observations}}
\tablehead{\colhead{Date}&\colhead{$\Delta t$}&\colhead{Frequency}&\colhead{Flux Density}&\colhead{Instrument}\\
\colhead{}&\colhead{(days)}&\colhead{(GHz)}&\colhead{($\mu$Jy)}&\colhead{}}
\startdata
2004 Apr. 13.11 & 380.63 & 4.86  &    631 $\pm$ 50 &  VLA-C  \\
2004 May. 29.12 & 426.64 & 4.86  &    516 $\pm$ 39 &  VLA-C  \\
2004 Jun. 01.15 & 429.67 & 4.86  &    485 $\pm$ 44 &  VLA-D  \\
2004 Jul. 19.94 & 478.46 & 4.86  &    441 $\pm$ 43 &  VLA-D  \\
2004 Aug. 15.79 & 505.31 & 4.86  &    528 $\pm$ 53 &  VLA-D  \\
2004 Sep. 03.79 & 524.31 & 4.86  &    331 $\pm$ 28 &  VLA-A  \\
2004 Dec. 09.66 & 621.18 & 4.86  &    361 $\pm$ 36 &  VLA-B  \\
2005 Jan. 29.38 & 671.90 & 4.86  &    346 $\pm$ 51 &  VLA-B  \\
2005 Apr. 07.21 & 739.73 & 4.86  &    292 $\pm$ 37 &  VLA-C  \\
2005 Jun. 12.78 & 806.30 & 4.99  &    240 $\pm$ 17 & EB+GBT+WB+AR+JB \\
\enddata
\tablenotetext{*}{NOTE - EB = 100m Effelsberg telescope; 
GBT = 105m Green Bank Telescope; WB = phased Westerbork array;
  AR = 305m Arecibo telescope; JB = 25m MarkII Telescope.}
\end{deluxetable}

\clearpage

\clearpage

\clearpage

\begin{figure}
\plotone{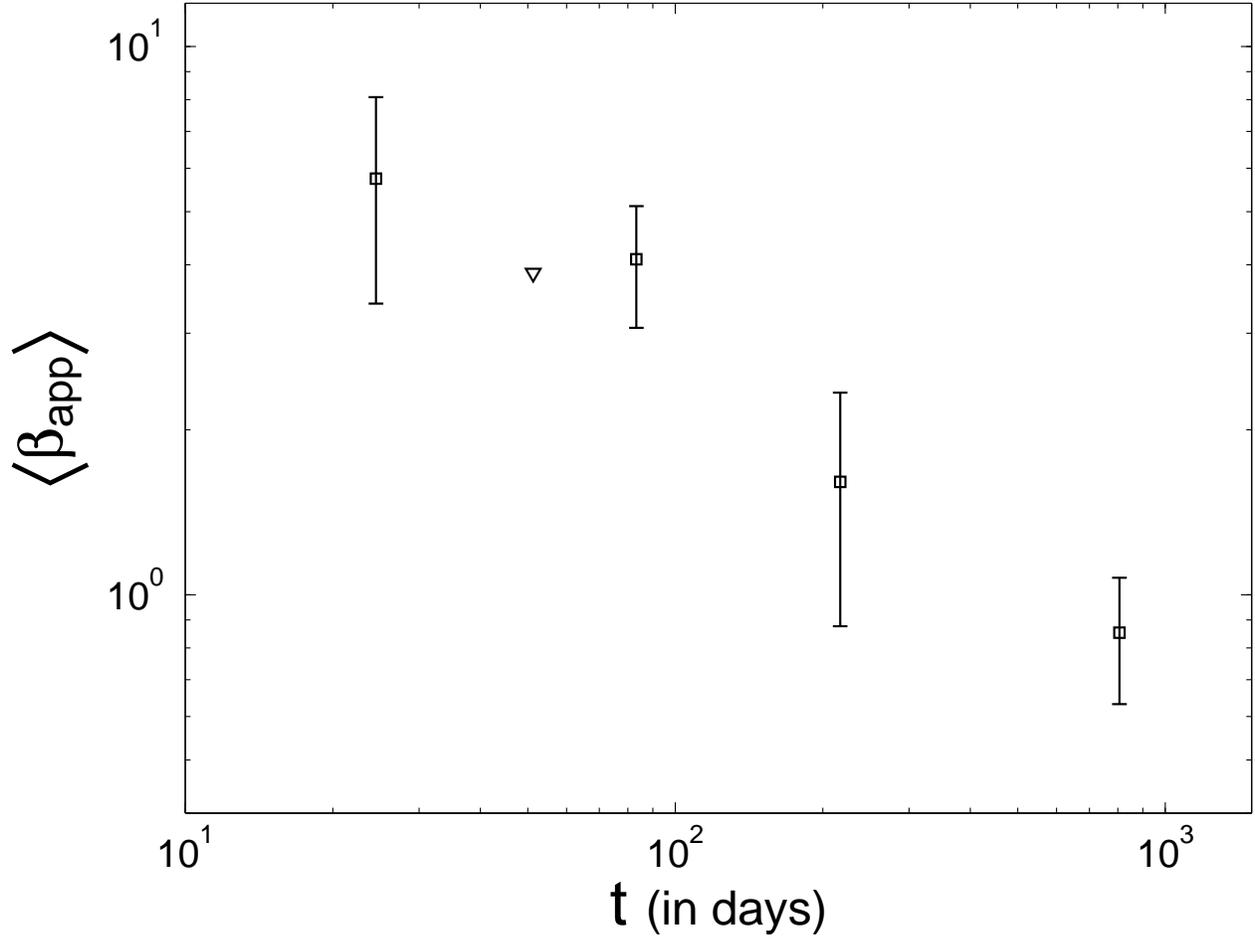}
\caption{ Evolution of the average apparent expansion velocity
derived from direct size measurements, and assuming a Gaussian
intrinsic surface brightness profile.
\label{FIG3}}
\end{figure}

\begin{figure}
\plotone{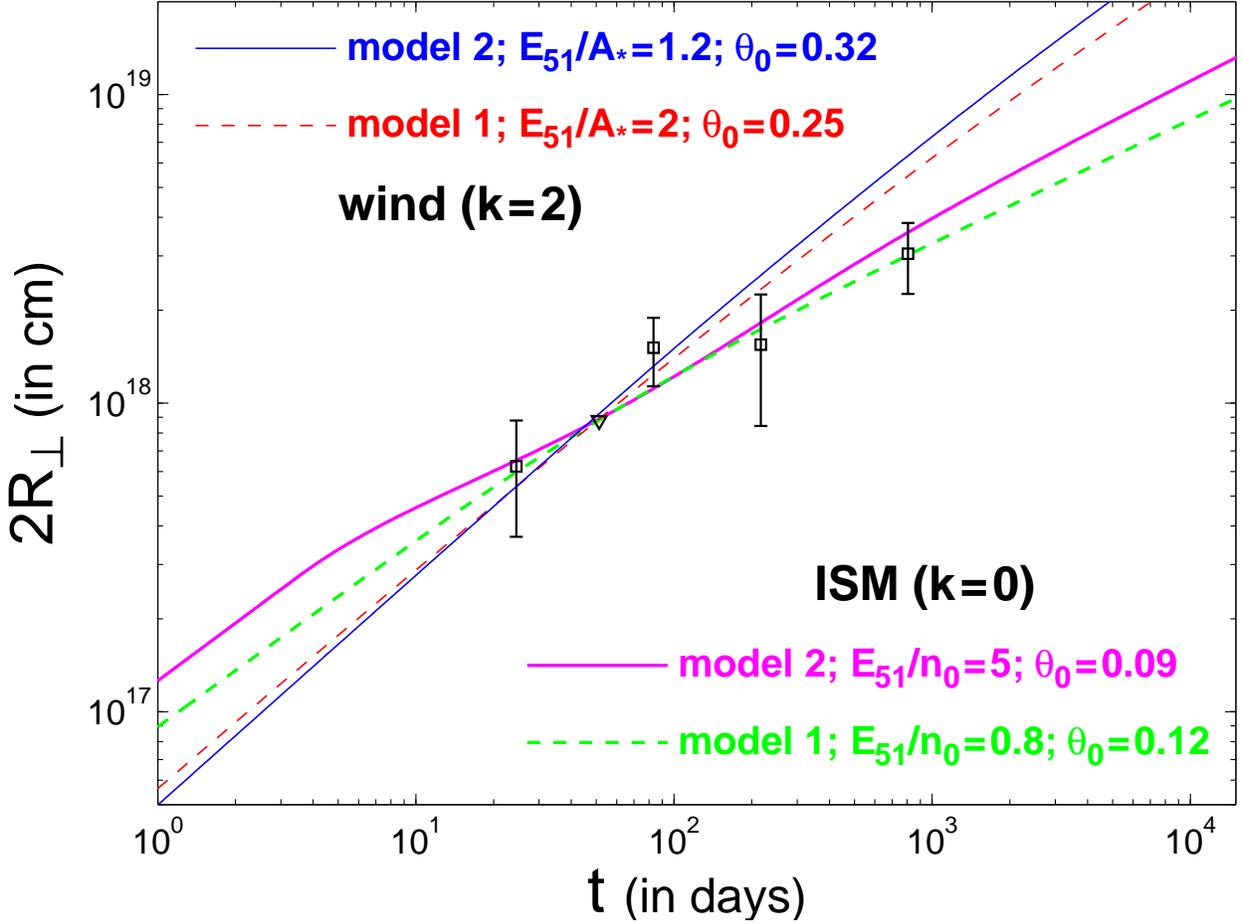} 
\caption{Tentative fits of theoretical models for the evolution of
the source size \citep[from][]{gra05} to the observed image size (of
diameter $2R_\perp$) of the radio afterglow of GRB~030329 up to
$83\;$days, with the addition of the measured sizes at $217\;$days
\citep[from][]{tay05}, and at $806\;$days (from this work). In model 1
there is relativistic lateral spreading of the GRB jet in its local
rest frame, while in model 2 there is no significant lateral expansion
until the jet becomes non-relativistic. The external density is taken
to be a power law with the distance $r$ from the source, $\rho_{\rm
ext} = Ar^{-k}$, where $k = 0$ for a uniform external density while $k
= 2$ is expected for a stellar wind environment.
\label{FIG4}}
\end{figure}

\clearpage

\begin{figure}
\plotone{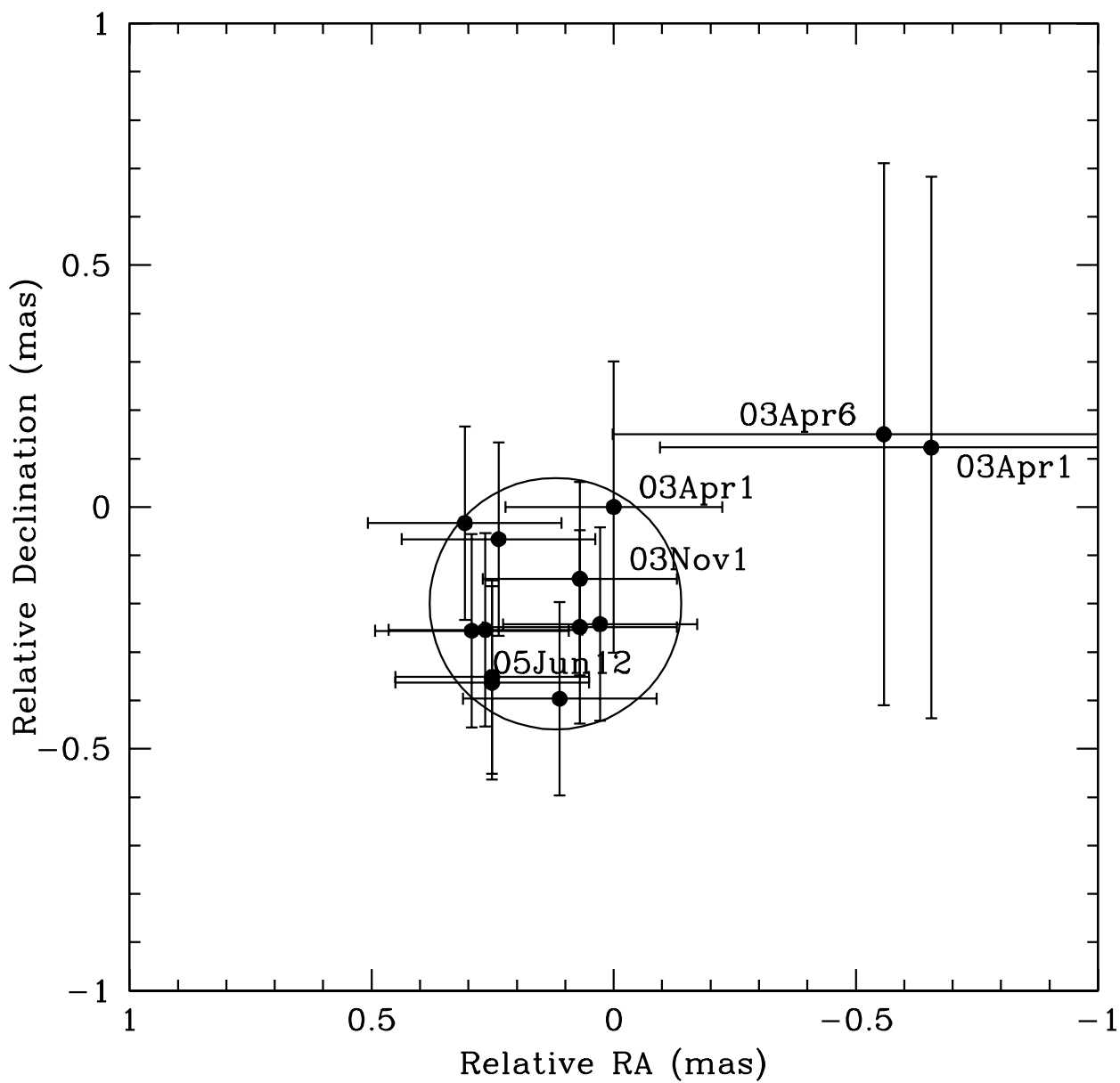}
\caption{The positions derived from the observations in eight
epochs relative to the first determination on April 1st at 8.4 GHz.
Observations at multiple frequencies at a given epoch have been
plotted separately since they are independent measurements.  A circle
with a radius of 0.26 mas (2$\;\sigma$) is shown to encompass all
measurements except those taken within the first eight days at 5 GHz, 
which suffer from systematic
errors \citep{tay04}.  Taken together these observations provide a
constraint on the proper motion of 0.35 $\pm$ 0.38 mas over 806 days.}
\label{skypos}
\end{figure}

\clearpage


\begin{thebibliography}{}

\bibitem[Berger {\it et al.}\ (2003)]{bkp+03}
Berger, E. {\it et al.}\  2003, Nature, 426, 154

\bibitem[Berger {\it et al.}(2001)]{BKF01}
Berger, E., Kulkarni, S.~R., \& Frail, D.~A. 2001, ApJ, 560, 652

\bibitem[Bloom, Frail \& Kulkarni(2003)]{BFK03}
Bloom, J. S., Frail, D. A., \& Kulkarni, S. R. 2003, ApJ, 594, 674

\bibitem[Chevalier \& Li(2000)]{cl00}
Chevalier, R.~A. and Li, Z. 2000, ApJ, 536, 195

\bibitem[Costa {\it et al.}(1997)]{cos97} Costa, E., et al.\ 1997, 
Nature, 387, 783 


\bibitem[Frail {\it et al.}\ (1997)]{fra97}
Frail, D.~A., Kulkarni, S.~R., Nicastro, S.~R., Feroci, M., and
  Taylor, G.~B. 1997, Nature, 389, 261

\bibitem[Frail et al.(2001)]{Frail01}
Frail, D. A., {\it et al.}, 2001, ApJ, 562, L55

\bibitem[Frail {\it et al.}(2005)]{fra05} Frail, D.~A., Soderberg, 
A.~M., Kulkarni, S.~R., Berger, E., Yost, S., Fox, D.~W., and Harrison, 
F.~A.\ 2005, \apj, 619, 994 

\bibitem[Gorosabel {\it et al.}(2006)]{gorosabel06} Gorosabel, J.,
  Castro-Tirado, A.J., Ramirez-Ruiz, E., Granot, J., Caon, N., Cairos,
  L.M., Rubio-Herrera, E., Guziy, S., de Ugarte Postigo, A., and
  Jelinek, M., 2006, ApJ, 641, L13

\bibitem[Granot(2006)]{Granot06}
Granot, J. 2006, to appear in Rev. Mex. Astron. Asrof. (astro-ph/0610379)

\bibitem[Granot(2005)]{granot05}
Granot, J.\ 2005, ApJ, 631, 102

\bibitem[Granot, Piran \& Sari(1999)]{gps99a}
Granot, J., Piran, T., and Sari, R. 1999, ApJ, 513, 679

\bibitem[{Granot} \& {Loeb}(2001)]{gl01}
Granot, J., and Loeb, A. 2001, ApJ, 551, L63

\bibitem[{Granot} \& {Loeb}(2003)]{gl03}
Granot, J. and Loeb, A. 2003, ApJ, 593, L81

\bibitem[Granot, Ramirez-Ruiz \& Loeb(2005)]{gra05} 
Granot, J., Ramirez-Ruiz, E., \& Loeb, A.\ 2005, ApJ, 618, 413

\bibitem[Granot, Nakar \& Piran(2003)]{gnp03}
Granot, J., Nakar, E., \& Piran, T.\ 2003, Nature, 436, 128

\bibitem[Granot \& Sari(2002)]{GS02}
Granot, J., \& Sari, R. 2002, ApJ, 568, 820

\bibitem[Greiner et al.(2003)]{gre03} Greiner, J., Peimbert, 
M., Estaban, C., Kaufer, A., Jaunsen, A., Smoke, J., Klose, S., \& Reimer, 
O.\ 2003, GRB Coordinates Network, 2020, 1 

\bibitem[Li \& Song(2004)]{li04} Li, Z.~\& Song, L.~M.\ 
2004, \apjl, 614, L17 

\bibitem[Mirabal \& Halpern(2006)]{mirabal06} Mirabel, N., \& Halpern,
 J.~P.\ 2006, GCN GRB Observation Report 7792

\bibitem[van Paradijs {\it et al.}(1997)]{van97} van Paradijs, J., 
et al.\ 1997, Nature, 386, 686 

\bibitem[Peterson \& Price(2003)]{pet03} Peterson, B.~A., \& 
Price, P.~A.\ 2003, GRB Coordinates Network, 1985, 1 

\bibitem[Piran(2005)]{Piran05}
Piran, T. 2005, Rev. Mod. Phys., 76, 1143

\bibitem[Piran, Nakar \& Granot(2004)]{png04} 
Piran, T., Nakar, E., \& Granot, J. 2004, in ``Gamma-Ray Bursts: 30
Years of Discovery'', ed.  E. E. Fenimore \& M. Galassi, AIP
Conference Proceedings, Vol.~727, p.~181

\bibitem[Resmi {\it et al.}(2005)]{resmi05}
Resmi, L., et al.\ 2005, A\&A, 440, 477

\bibitem[Sari, Piran \& Narayan(1998)]{SPN98}
Sari, R., Piran, T., \& Narayan, R. 1998, ApJ, 497, L17

\bibitem[Sheth {\it et al.}(2003)]{she03} Sheth, K., Frail, D.~A., 
White, S., Das, M., Bertoldi, F., Walter, F., Kulkarni, S.~R., \& Berger, 
E.\ 2003, \apjl, 595, L33 

\bibitem[Taylor {\it et al.}(1999)]{tay99} Taylor, G.~B., Carilli,
C.~L., \& Perley, R.~A.\ 1999, ASP Conf.~Ser.~180: Synthesis Imaging in
Radio Astronomy II, 180.

\bibitem[Taylor {\it et al.}(2004)]{tay04}
Taylor, G.~B., Frail, D.~A., Berger, E., and Kulkarni, S.~R. 2004, ApJ,
  609, L1

\bibitem[Taylor {\it et al.}(2005)]{tay05} 
Taylor, G.~B., Momjian, E., Pihlstr\"om, Y., Ghosh, T., \& Salter, C.
2005, ApJ, 622, 986

\bibitem[Vanderspek et al.(2003)]{van03} Vanderspek, R., et 
al.\ 2003, GRB Coordinates Network, 1997, 1 

\bibitem[van der Horst et al.(2005)]{vanderhorst05} van der Horst,
A.~J., Rol, E., Wijers, R.~A.~M.~J., Strom, R., Kaper, L., \& Kouveliotou,
C.\ 2005, \apj, 634, 1166


\end{thebibliography}
\end{document}